\documentclass[12pt,cites,graphicx,a4paper]{article}
\usepackage{graphicx}
\usepackage{epsfig}
\usepackage{amssymb}

\begin{document}

\title{ On the predictive power of database classifiers formed by a small network of interacting chemical oscillators. }

\author{L. Zommer$^{1}$, K. Gizynski$^{1}$ and J. Gorecki$^{1,*}$\\
$^{1}$Institute of Physical Chemistry, Polish Academy of
Sciences,\\ Kasprzaka 44/52, 01-224 Warsaw, Poland.\\
$^{*}$jgorecki@ichf.edu.pl\\
}

\maketitle

\begin{abstract}
{\it The predictive ability of database classifiers constructed with a network of interacting chemical oscillators is studied. Databases considered here are composed of records, where each record contains a number of parameters (predictors) characterizing the case and the output variable describing the case type. In a series of recent papers we have discussed the top-down design of database classifiers, that return output information as the number of excitations observed at a selected droplet of the network. The design is based on an evolutionary algorithm that optimizes the network to achieve maximum mutual information between the network evolution and the record types. Here we discuss results illustrating that such classifiers do have a predicting ability. They are able to give a correct classification for database records that are not included into the process of classifier training. As example  databases we consider points inside a multidimensional unit cube and test if they belong to a ball located at the mid of cube. }
\end{abstract}



\section{Introduction}

Silicon technology has dominated information processing applications for over 50 years. It is anticipated that the Moore law \cite{moore} describing its progress will finally break down and the rate of progress will slow. Therefore, there is a growing interest in alternative information processing strategies: quantum, optical or these based on chemical reactions in hope to support the progress 
or even lead to a leap in the rate of information processing \cite{b1,b2,b3,b4,b5}. 

Among the considered information processing media the systems using chemical reactions attract significant scientific attention. For example, the interest in reaction-diffusion nonlinear chemical media and in information processing with propagating pulses of excitation \cite{b4} have been motivated by similarity to signals in nerve systems of living organisms \cite{haken}.
It is expected that deep understanding of chemical information processing will allow to implement efficient algorithms for navigation or image recognition that are executed by animal brains.

There is a huge difference in technologies for information processing using semiconductors and the one based on chemical reactions. After many years of progress we know how to fabricate small, but extremely reliable logic gates. The time of their stable functioning is measured in years. Having logic gates we can use a bottom-up approach and assemble them into more complex information processing structures, like for example a processor \cite{feynm}. On the other hand, the technology for information processing based on chemical reaction-diffusion processes is at very early stage of development. For example, the area of information processing region of a logic gate based on Belousov-Zhabotinsky (BZ) reaction and using information coding with pulses of excitation is circa $1 cm^2$ . In typical experiments \cite{yoshi-2009} such a gate reliably operates for about $30$ minutes. Such functioning time is sufficient to perform a few operations and demonstrate that the device works, but it is too short to construct a complex circuit.
Therefore, the bottom up design from gates to an universal computer that works perfectly with semiconductors seems non-applicable for information processing devices that use a nonlinear chemical medium.

A few years ago a new approach to chemical computing was proposed \cite{dittrich}. 
It has been demonstrated that many information processing tasks can be performed by networks of interacting chemical oscillators using their specific properties.
During a single oscillation cycle of a typical chemical oscillator we can distinguish three phases: excited, refractive and responsive \cite{d1,d2,cmst}.
Such distinction is important for the simplified model of interactions between two oscillators coupled by the exchange of reaction activator. The excited phase denotes the peak of activator concentration. An excited oscillator is able to spread out activator molecules and to speed up their production  in the medium around. In the refractory phase the concentration of inhibitor is high and in this phase the oscillator does not respond to activator transported from neighboring oscillators. In the responsive phase the concentration of the inhibitor decreases. An oscillator  in  this phase can get excited by interactions with an oscillator in the excited phase.

For information processing it is crucial that  oscillators can be individually controlled by an external factor. In the case of a photosensitive variant of BZ reaction it is illumination with a blue light that activates Ru-complex and leads to production of reaction inhibitor\cite{photosens-BZ}, so in the following we use this name illumination to describe the controlling factor. For the analysis presented below it is sufficient to assume that the controlling factor has an inhibiting effect. When the controlling factor is switched on oscillations are suppressed, when it is off oscillations are restored \cite{ pccp-3drop}. The control factor can be also used to input information into the network as described later.

In experiments a network of oscillators, working as a dataset classifier can be made of droplets containing oscillatory BZ-medium. Such droplets, immersed in a solution of lipids in hydrocarbons are stable and can touch each other without coalescence \cite{szyman}. The touching droplets can communicate via exchange of the activator. Having this analogy in mind we use the term droplet in the following text to refer to individual oscillators of the network. An alternative chemical network that performs similar function can be formed by closely located dark regions on an illuminated membrane containing reagents of a photosensitive BZ reaction \cite{adamatz-PRE}. 

In the considered classification problems the database was represented by a set of records.
Each record had the form of an $(n+1)-$tuple $(p_1, p_2,..., p_n, rt )$ where $p_i$ are predictors 
represented by real or integer numbers and $rt$ is an integer defining the record type.
For example the Wisconsin Breast Cancer Dataset \cite{WBCD} contains records where predictors describe physical properties of a cancer cell and the record type indicates the form of cancer (malignant or benign). 
An algorithm that solves the dataset classification problem is supposed to return the record type
after the values of predictors are given as the inputs.

In our recent publications based on computer simulations \cite{d1,d2,cmst,cancer} we postulated that networks of interacting chemical oscillators can solve dataset classification problems with a reasonable accuracy. Here we present results for classifiers formed by $9$ oscillators arranged on the nodes of a hexagonal lattice as illustrated in Fig. 1. At the beginning we have to fix the time interval $[0, t_{max}]$ within which the time evolution of the network  is observed. This is an important assumption, because it indicates that information processing is a transient phenomenon. We assume that the output information can be extracted by observing the system within the time interval $[0, t_{max}]$ and reaching a steady state at a long time is not important for the network answer.
We also assume that there are two types of oscillators in the network: the normal ones or the input ones. The state of each oscillator in the network is controlled by illumination. For the normal droplet  $k$ oscillations are inhibited within the time interval $[0,t^{(k)}_{illum}]$, and $ t^{(k)}_{illum}$ ($0 \le t^{(k)}_{illum} \le t_{max}$) is the time at which the oscillation cycle of the $k$-th droplet is restored. For a given classification problem all times $ t^{(k)}_{illum}$  are the same for all processed records and they make a "program" executed by the network. If an oscillator is the input for the $j$-th predictor, and if the predictor value is $p_j$ then this oscillator is inhibited (illuminated) within the time interval $[0, A*p_j+B]$. The values of $A$ and $B$ depend on the classification problem, but they are the same for all predictors. Illumination of all normal droplets is fixed therefore, for a selected database record inhibitions of all oscillators in the network are known. Now we can observe the network evolution in time. Here we assume that the number of oscillations of a selected oscillator represents the network output. The output droplet is selected such that for this droplet we obtain maximum mutual information between the record types of the training dataset used for network optimization and the number of droplet oscillations.

Of course, it would be naive to believe that a randomly selected network of oscillators performs a 
desired function. The network parameters, such as $A, B$ or $\{ t^{(k)}_{illum} \}$ should be optimized according to the problem we are going to solve. To this end we select a training database of the problem and perform a complex, multi-parameter optimization of the network using an evolutionary algorithm. The details of the method are briefly described in the next Section. Our recent papers have demonstrated \cite{d1,d2,cmst,cancer}  that after optimization a classifier based on coupled oscillators can be trained to perform its function with quite high accuracy. Moreover, it has been discovered that optimized classifiers do have a predicting ability. For the Wisconsin Breast Cancer Dataset a classifier trained on a part of the original dataset of the problem can correctly determine the cases not included during the training with accuracy of over $88 \%$ \cite{cancer}. However, the Wisconsin Breast Cancer Dataset is quite small and contains only $699$ records. In the aforementioned study we separated the database into two disjoint datasets: $599$ records were used for system training and the remaining $100$ formed the test dataset. However, it may be expected that the test dataset of the problem was too small for the objective evaluation of classifier predictive power.

Here we investigate classifier predictive ability considering another problem, for which datasets of an arbitrary size can be easily generated. In our previous paper \cite{cmst} we demonstrated, using computer simulations, that a regular network of 16 droplets can determine with a reasonably high accuracy (over $80 \%$) if three random numbers in the range $[0,1]$ describe a point located inside a ball inscribed within the cube $[0,1] \times [0,1] \times [0,1]$ or not.
Here we follow this geometrically oriented  problem and consider location of a point from a unit cube with respect to a ball in a multi-dimensional space.

Let us define the n-dimensional unit cube:\\
$$ Q_n=\overbrace{ [0,1] \times [0,1] \times \cdots \times [0,1]}^{n\ \rm times},$$ where $n \in \{3,4,5,6\}$ \\
and\\
the n - dimensional ball $ S_n(a_0, r)$, characterized by the center $a_0 = (0.5, 0.5, ..., 0.5)$ and the radius $ r$:\\
$$ S_n( a_0, r) = \{p \in {\bf R}^n; || p - a0 || \le r \}$$
where the symbols $|| .||$ denote the Euclidian distance.

We consider a database in which predictors represent coordinates of a point $p$ in the cube $ Q_n$ and the record type value of $ rt $ is $1$ if a point belongs to the ball and $0$ if it does not.

In 3-dimensions the volume of a ball inscribed in a cube is approximately $1/2$ of cube volume, but in higher dimensions the ball volume decreases. Therefore, if we consider an inscribed ball then in higher dimensions the database is dominated by points located outside the ball. To get maximum entropy of the dataset record types, equal to 1bit, we select the radius $ r_n$ such that:\\
$V ( Q_n \cap S_n( a_0, r_n)) / V (Q_n) = 1/2 $.\\
where $V(Z)$ denotes n-dimensional volume of the set $Z$.
The condition is satisfied when 
for: $ r_3=0.49175, r_4=0.5685, r_5=0.6395, r_6=0.7015$ where the index denote system dimension.

In this paper we optimize the networks illustrated in Fig.1 to classify the Ball-in-Cube problem defined above. We use two different sizes of the training datasets: $500$ or $5000$ records and different lengths of optimization. Next, the optimized classifiers are evaluated on much larger test datasets to verify if they can correctly determine types of records excluded from the training. On the basis of classification accuracy at the testing datasets we conclude on the classifier predictive ability.

The paper is arranged into three sections. The first one contains the basic information on the model of oscillations and the genetic algorithm used for classifier optimization. A reader who is not familiar with the subject is advised to read our open access paper \cite{cmst} where the details of the methods are given. In the next Section we present results for classifiers of the Ball-in-Cube problem for different dimensions of space. In the final Section we discuss the predictive ability of the considered classifiers.

\begin{figure}
\begin{center}
\scalebox{0.7}{\includegraphics{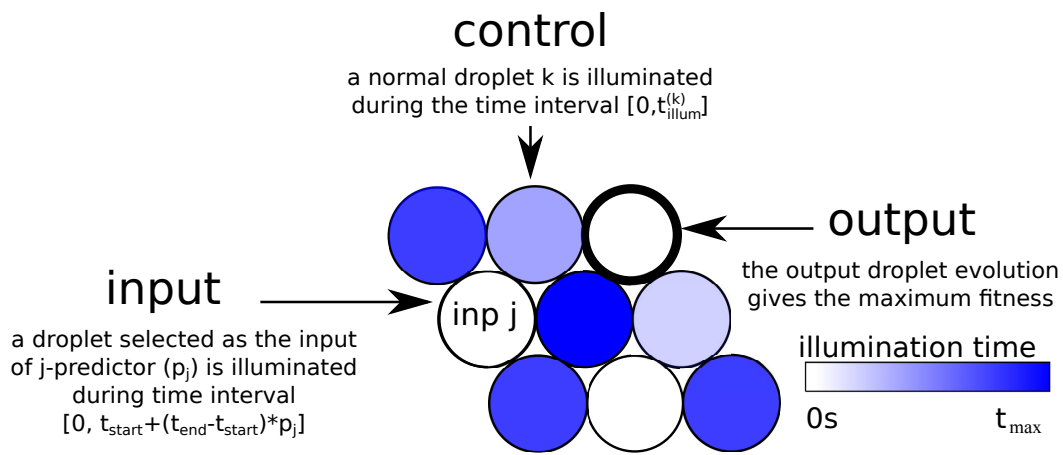}}
\end{center}
\caption{ Illustration of geometry of a network of oscillators (droplets) considered in the paper. The network is formed of two types of droplets: "normal" ones for which the inhibition time is the same for all records and the "input" ones for which the time of inhibition depends on the value of a corresponding predictor in the processed record. The number of oscillations of a selected output droplet within the time interval $[0, t_{max}]$ (here $ t_{max} =400$ $ t.u.$) is regarded as the network output. Here, as well as in the other figures the intensity of blue color represents illumination time for a particular normal droplet and the output droplet is marked by a thick border. }
\label{fig-1}
\end{figure}

\clearpage

\section{ Evolutionary optimization of a classifier} 

In this Section we briefly describe the methods used to optimize the network of chemical oscillators to perform a selected classification task. An extended description of the optimization algorithm can be found in our open access paper \cite{cmst}.

A classifier $C$, like the one shown in Fig.1, can be formally defined as an 7-tuple:\\
$$ C = \{ N, G, t_{max}, S, \{ t^{(k)}_{illum}, k=1,N \}, A, B \}$$
where: $ N$ stands for the number of oscillators involved, G describes the network geometry and interactions between oscillators, 
$t_{max}$ defines the end of time interval within which the time evolution of the network  is observed, S is the structure 
of droplet types, $\{ t^{(k)}_{illum}, k=1,N \}$ is the set of illumination times for normal droplets and parameters $A, B \ge 0$ define the affine function that translates the predictor value into illumination of the corresponding input droplet. We assume that predictor values are normalized to the interval $[0,1]$. If the predictor value is $0$ then the illumination time is the shortest and equals $t_{start} = B$. On the other hand if the predictor value is $1$ then the illumination time is the longest and equals $t_{end} = A+B$. Our optimization program searches for the best values of $t_{start}$ and $t_{end}$ rather than for $A$ and $B$.

In the following analysis we fix $N =9$, the network has a hexagonal geometry with interactions between the nearest droplets and $ t_{max} =400$ $ t. u.$. Such network is defined by parameters: $\{S, \{ t^{(k)}_{illum},  k=1,N  \}, t_{end}, t_{start} \}$. We perform optimization with respect to all these parameters. The problem is very complex thus we apply evolutionary optimization to obtain the best classifier of the training dataset. We believe that if the training dataset objectively reflects the data structure of the problem then the network optimized for correct classification of the training dataset will also be useful for other datasets of a given problem. 

We assume that the number of excitations of a selected droplet is the classifier output. Such choice is simple, but arbitrary. One can consider any other translation of network evolution into the output string. 
The quality of classifier is measured as the maximum mutual information between the record types of the training dataset used for network optimization and the network outputs. The mutual information between two sets \cite{infor} is the quantity that describes how much information about an element of one set can be gained if we know the element of the other. Let us assume that the training dataset is D and consider its classifier C. For each record $r$ ($r \in D$) characterized by the record type $rt$ we can detect the number of excitations $o_{rj}$ observed at the droplet $j$ of the network. For the classifier C we select the output droplet as the one for which the mutual information
between the sets $R=\{ rt, r \in D \}$ and $O_j=\{ o_{rj}, r \in D \}$ has the maximum value. The value $I(R,O)$:\\
$$I(R,O) = max_j (H(R) + H(O_j) - H(R, O_j) $$
is considered as the quality of a classifier and it is used as the fitness function during optimization of classifiers parameters.
In the formula above $H(R)$ and $H(O_j)$ are the Shannon entropies \cite{shan} of the set of output types in the training dataset and of the set of numbers of excitations observed at the droplet $j$ of the network whereas $H(R, O_j)$ is the joint entropy of both these sets. In order to calculate $I(R,O)$ we have to study the network evolution on all elements of the training dataset. It means that the training dataset should not be too large and, moreover, that we need a fast algorithm to describe the network time evolution including possible interaction between oscillators.

Following the previous papers we used event-based-model that reflects basic properties of interacting droplets containing BZ-medium \cite{cmst}. The parameters of the model used in calculations presented below has been adjusted to reflect our recent experiments  with the networks \cite{kg-priv}. We assume that during the oscillation cycle a droplet can be in one of there phases: the excitation phase lasting 0.01 time unit, the refractory phase, lasting 34 time units or the responsive phase that is 50 time units long. After the responsive phase the excitation phase follows and the cycle repeats. Thus, the period of the cycle is $84.01$ time unit. The separation of oscillation cycle into phases allows to introduce a simple model for interaction between droplets. A droplet in the responsive phase can be excited by an excited nearest neighbor. A droplet in the refractory phase does not response to such external excitations. On the other hand, if a droplet is in the responsive phase then and excited neighbor can change its time evolution. We assume that if a droplet is excited then 9 time units later all its neighboring droplets that are in the responsive phase switch into the excited phase. We also assume that if a droplet is illuminated then it remains in the refractory phase for 20 time units after the illumination is switched off and next the responsive phase starts. It means that after illumination is switched off the first excitation appears after 70 time units (20 time units of the refractive phase plus 50 time units of the responsive phase) unless the droplet during the responsive phase gets excited by its neighbor.

Because the manual programming of nonlinear media is hard and time consuming, we introduce an evolutionary algorithm to
find the optimum network structure and the set of illumination times. The optimization algorithm is initialized with a population of $500$ randomly generated classifiers. The fitness of each classifier is evaluated and the bottom $40 \%$ of individuals are discarded. The top $10 \%$ of classifiers are copied to the next generation. The remaining $90 \%$ ($450$)  individuals of the next generation are obtained by recombination and mutations within the upper $60 \%$ 
of classifiers in the previous generation. The operations are illustrated in Fig. 2. 
Following the standard recombination techniques 
\cite{Goldberg1989} $ 2$ parents are involved in the procreation of one offspring. A parallelogram shaped sub-grid of 
droplets from the Parent 1 constrained by two, randomly selected points A and B replaces the corresponding sub-grid 
in the Parent 2 as illustrated in Fig. 2 to yield a new individual. The illumination interval of 
input droplets co-evolved with the network is copied from the Parent 1 to the Offspring. 
The newly generated individual 
is subjected to three, subsequent mutation operators with fixed mutation rates executed in the following order:
\begin{enumerate}

\item {Input illumination interval mutation}

Times determining illumination interval of droplet with input types, i.e. $t_{start}^\mathrm{(Offspring)}$, 
$t_{end}^\mathrm{(Offspring)}$ are selected from the normal distributions with the averages 
$t_{start}^\mathrm{(Parent1)}$, $t_{end}^\mathrm{(Parent1)}$ respectively and $\sigma^2=10$ as follows: 
$$
t_{start}^\mathrm{(Offspring)}=min( \mathcal{N}(t_{start}^\mathrm{(Parent1)},\sigma^2), \mathcal{N}(t_{end}^\mathrm{(Parent1)},\sigma^2) ) 
$$
$$t_{end}^\mathrm{(Offspring)}=max( \mathcal{N}(t_{start}^\mathrm{(Parent1)},\sigma^2), \mathcal{N}(t_{end}^\mathrm{(Parent1)},\sigma^2) ) $$

where $\mathcal{N}(t,\sigma^2)$ is a random number selected from the normal distribution with average $t$ and variance 
$\sigma^2$.

\item {Droplet type mutation}

We assume that input droplets can change into normal droplets and vice versa. The probability of droplet transformation 
is $p_{type}=0.04$. Regardless of droplet type the probability of obtaining an input droplet 
is $p_{inp}=0.12$ and for the normal one $1-p_{inp}$.

\item {Illumination time mutation}

If a droplet $k$ is of the normal type, then with a probability $p_{illum} = 0.04$ its illumination time is mutated.
New illumination time for the Offspring $t^{(k)}_{illum}$ is generated from the normal distribution with 
the average $t^{(k)}_{illum}$ and $\sigma^2 = 25$. 
\end{enumerate}

The values of all probabilities were selected arbitrary as being reasonably small, but still large enough to produce noticeable 
changes in network functionality after mutation. They should have little influence on the final, optimized classifier, 
but definitely regulate the rate of convergence towards the optimum solution.

In the second stage of mutation procedure the droplet types are also subjected to evolution and thus the position of the inputs might vary 
according to generation. We also do not introduce any constraints on the number of input droplets in the network. 
For example it is possible that there are a few input droplets of one predictor and no input droplets of another. In this situation no information about the missing predictor is transferred to the network.

Exactly one droplet of the network is selected as the output droplet. The mutual information with the  
distribution of record types  is checked separately for each droplet in the network during the fitness evaluation procedure and 
the one with the highest value is selected as the output droplet. Since the mutual information in the droplets 
changes during evolution, the position of the output is not fixed and also can change from generation to generation. 
We do not exclude the case in which an input droplet is used as the output one.

\clearpage

\begin{figure}
\begin{center}
\scalebox{0.6}{\includegraphics{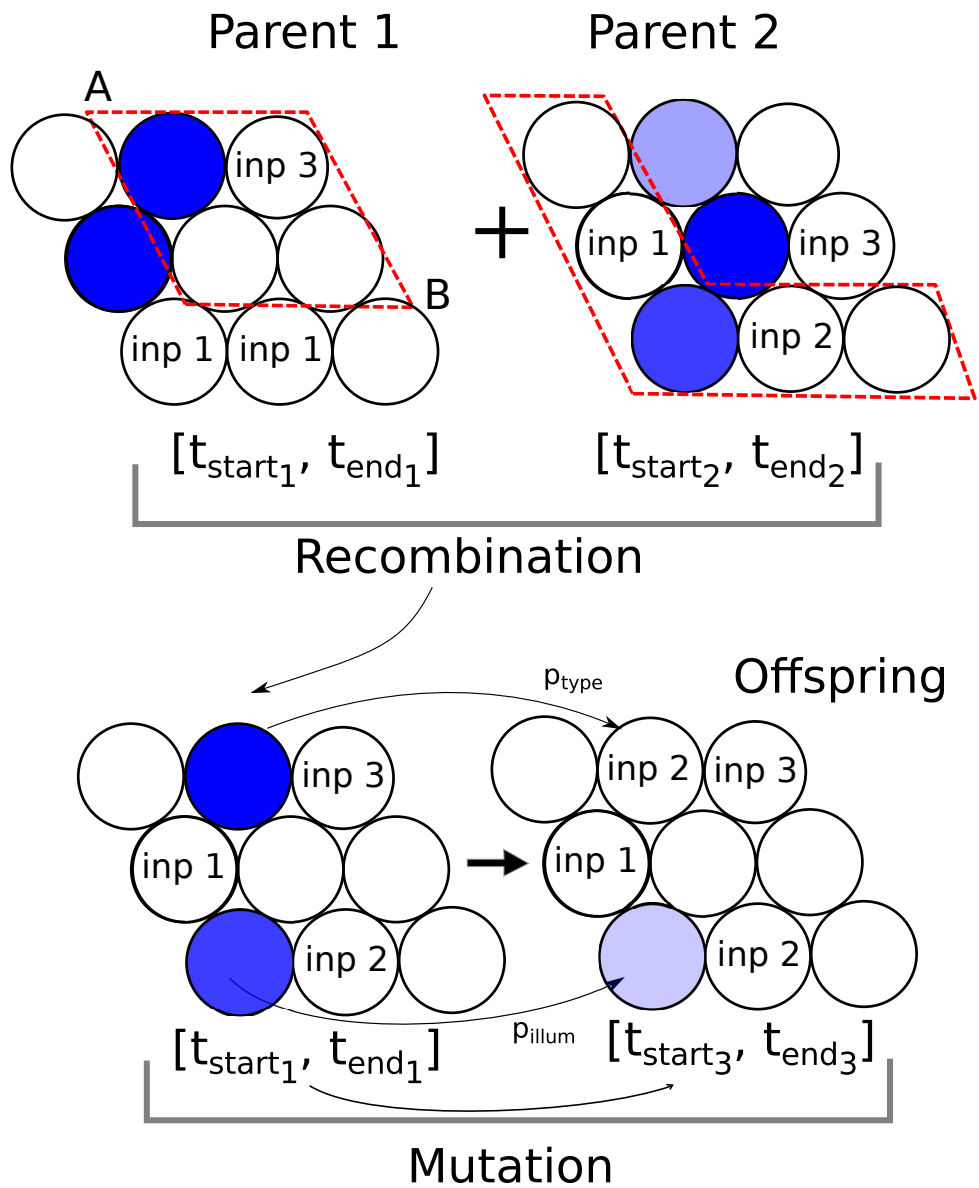}}
\end{center}
\caption{
Randomly selected points A and B in the structure of Parent 1 mark a parallelogram, which is copied, along with 
the input  illumination parameters, to the Offspring during the recombination process. The other part of 
the Offspring comes from the Parent 2. Then, during the mutation, droplet types 
and initial illumination times are modified. Intensity of blue color in each droplet is proportional to its 
illumination time. 
}
\label{fig-2}
\end{figure}
\clearpage

\section{Results}

We performed two types of optimization for the Ball-in-Cube problem: a short one using a small training dataset ( $500$ generations and $500$ records, named bu500 in the following text) and a long one based on a larger dataset ( $5000$ generations and $5000$ records, named bu5000). In the optimization algorithm we always worked with population of $500$ classifiers. After optimization was completed the quality of the obtained classifiers were verified on test databases with different sizes: $200, 500, 1000, 5000, 20000$ and $100000$ records. The predictor values of the database records were generated in a single program using a uniform random number generator in the range $[0,1]$, so we believe there is no overlap between the records. The record type was represented by $1$ if the predictors described a point belonging to the ball $ S_n(a_0, r)$ and $0$ for a point located outside the ball. 

The results obtained after long optimization of a classifier are shown in Figures from 3 to 6 and the figure number corresponds to the dimension of space for which the problem was studied. In all these figures the structure of subfigures is the same. 

Figures (a) show the fitness as the function of generation number. In all cases the major increase in fitness is observed within the first few hundred generations. After the rapid initial increase the fitness remains almost constant ($ n = 3$ and $ n = 6$) or reaches a metastable plateau within $1500$ generations ($ n = 4$ and $ n = 5$) . The maximum fitness (cf. Fig. 7a) is the decreasing function of space dimension. It can be explained by increasing complexity of the classification problem, whereas $t_{max}$ and the number of oscillators involved in classification remains unchanged. For 3-dimensional case the best classifier made of 9 droplets in the hexagonal geometry was characterized by fitness of $\approx 0.53$ bit. It is much more than obtained for the inscribed ball problem classified using the network of the same size formed on a regular lattice ($\approx 0.40$ bit for the training dataset of 200 records and $\approx 0.34$ bit for the training database of 400 records \cite{cmst}). It suggests that for a fixed number of oscillators a hexagonal geometry of droplets makes a better classifier. On the other hand the problem studied in \cite{cmst} was slightly different because the inscribed ball radius is by $ 2\%$ larger than the ball considered in this paper and the parameters of event-based-model used in \cite{cmst} were different.

Figures (b) illustrate the location of the input and output droplets and the parameters for the networks evolved with bu5000 strategy. The intensity of blue color reflects the illumination time for normal droplets $t^{(k)}_{illum}$. Each droplet represents a pie chart in which the arc length of red slice, normalized to the circumference gives the mutual information between the number of droplet excitations and record types measured in bits. In most of cases input droplets representing all predictors are directly interacting with the output one. The only exception are results for $n=4$, where the input of 4-th predictor is separated from the output. We believe that in this case optimization has not reached the global maximum and the contact will be established after a longer run.

Figures (c) present the probability distribution of different number of excitations at the output droplet as the function of the record type for bu5000. The yellow bars refer to points inside a ball and the green bars to the points outside. These results allow us to find a classification rule for the space dimensions $ n \in \{3, 4, 5\} $: if there are 3 or less excitations then the predictors describe a point inside the ball, otherwise the point is located outside. A similar rule, but with the threshold at the level of 2 excitations applies to the 6-dimensional problem.

Finally the Figures (d) show the accuracy of classifiers for test datasets of different sizes. We compare results for classifiers optimized using bu500 (the open symbols) and bu5000 (the filled symbols). The classification rules were derived on the basis of probability distributions for different numbers of excitations at the output droplet as functions of the record type observed for the corresponding training datasets (for bu5000 shown in Figures (c)). It can be seen that in most of cases the classifier accuracy on the training dataset is much higher than on the test databases (filled and open triangles for bu5000 and bu500 respectively). The differences between results for test databases containing $20000$ and $100000$ records were small, thus we claim that the accuracy observed for the largest test dataset is an objective measure of classifier quality. For 3-dimensional case the accuracy of optimized classifier based on hexagonal network ($88 \%$) is higher than the one obtained on the training dataset for the 9-droplet regular network ($\approx 84 \%$ for the training dataset of 200 records and $\approx 79 \%$ for the training dataset of 400 records) \cite{cmst}.

\clearpage

\begin{figure}
\begin{center}
\scalebox{0.6}{\includegraphics{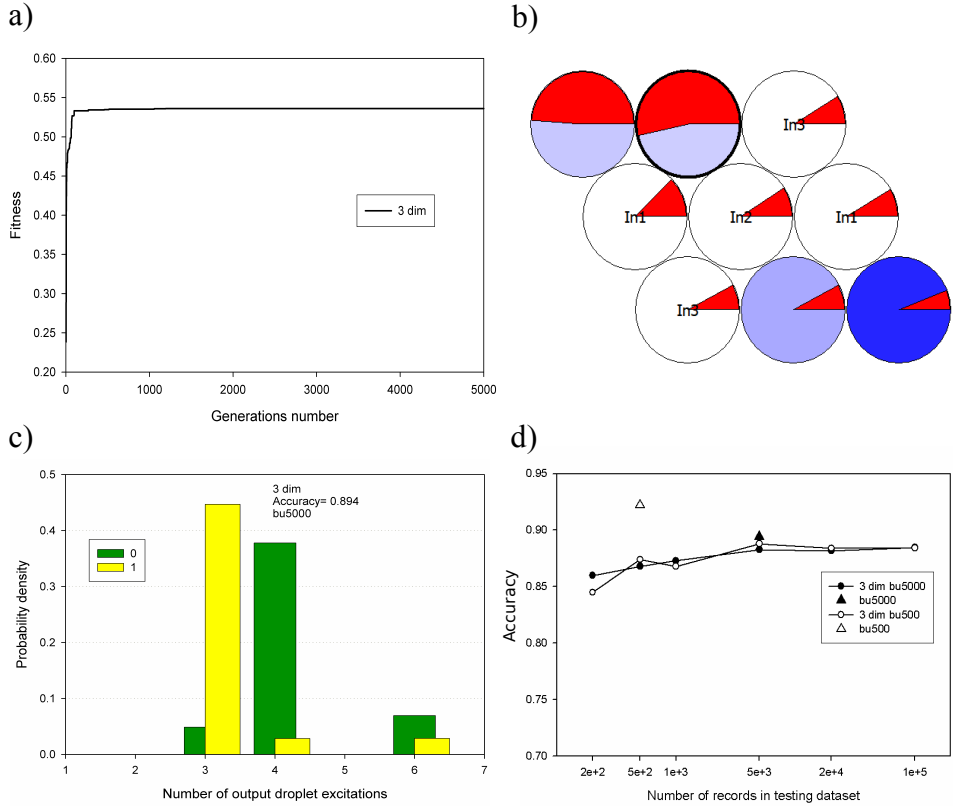}}
\end{center}
\caption{ The optimization of a classifier for Ball-in-Cube problem in 3 dimensions. Results obtained for population of $500$ individuals. Figures (a-c) illustrate results of bu5000 optimization, figure (d) compares bu500 and bu5000.
(a) The mutual information between the dataset types and the number of oscillations of the output droplet as the function of optimization progress.
(b) The structure of the best classifier. The arc length of red slice, normalized to the circumference gives the mutual information between the number of droplet excitations and record types measured in bits. 
(c) The probability density of different number of excitations at the output droplet of the optimized classifier observed for 
the training dataset of $5 000$ records. The yellow 
bars mark points inside the ball, the green ones mark the points outside. From this result we derive the classification rule.
(d) The accuracy of the best classifier on the test datasets of different size. The open and filled symbols mark results obtained for bu500 and bu5000 respectively. Triangles correspond to the accuracy obtained for the training datasets.
}
\label{fig-3}
\end{figure}

\clearpage

\begin{figure}
\begin{center}
\scalebox{0.7}{\includegraphics{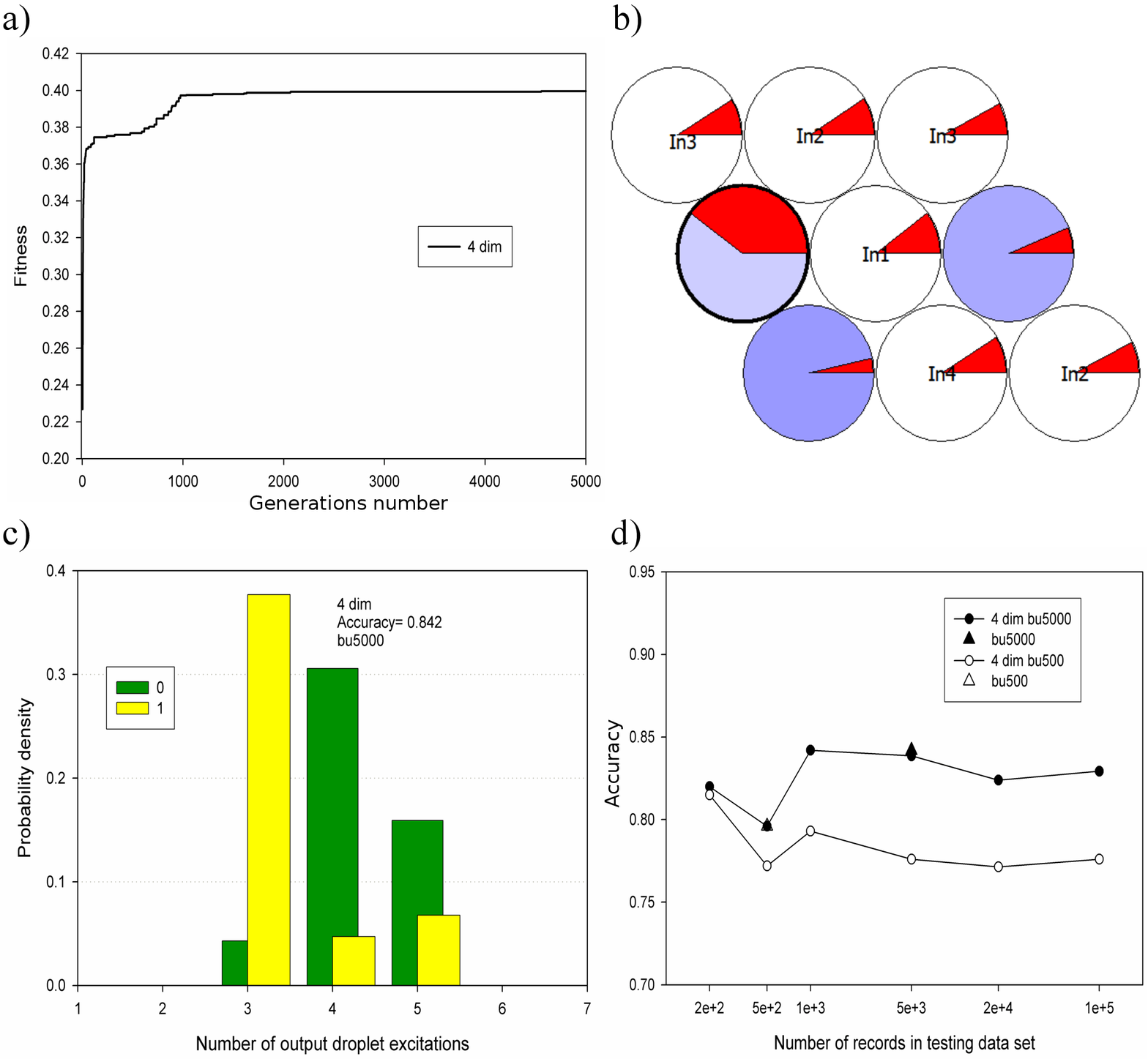}}
\end{center}
\caption{ The optimization of classifier for Ball-in-Cube problem in 4 dimensions. Notation as in Fig. 3. }
\label{fig-4}
\end{figure}

\clearpage

\begin{figure}
\begin{center}
\scalebox{0.7}{\includegraphics{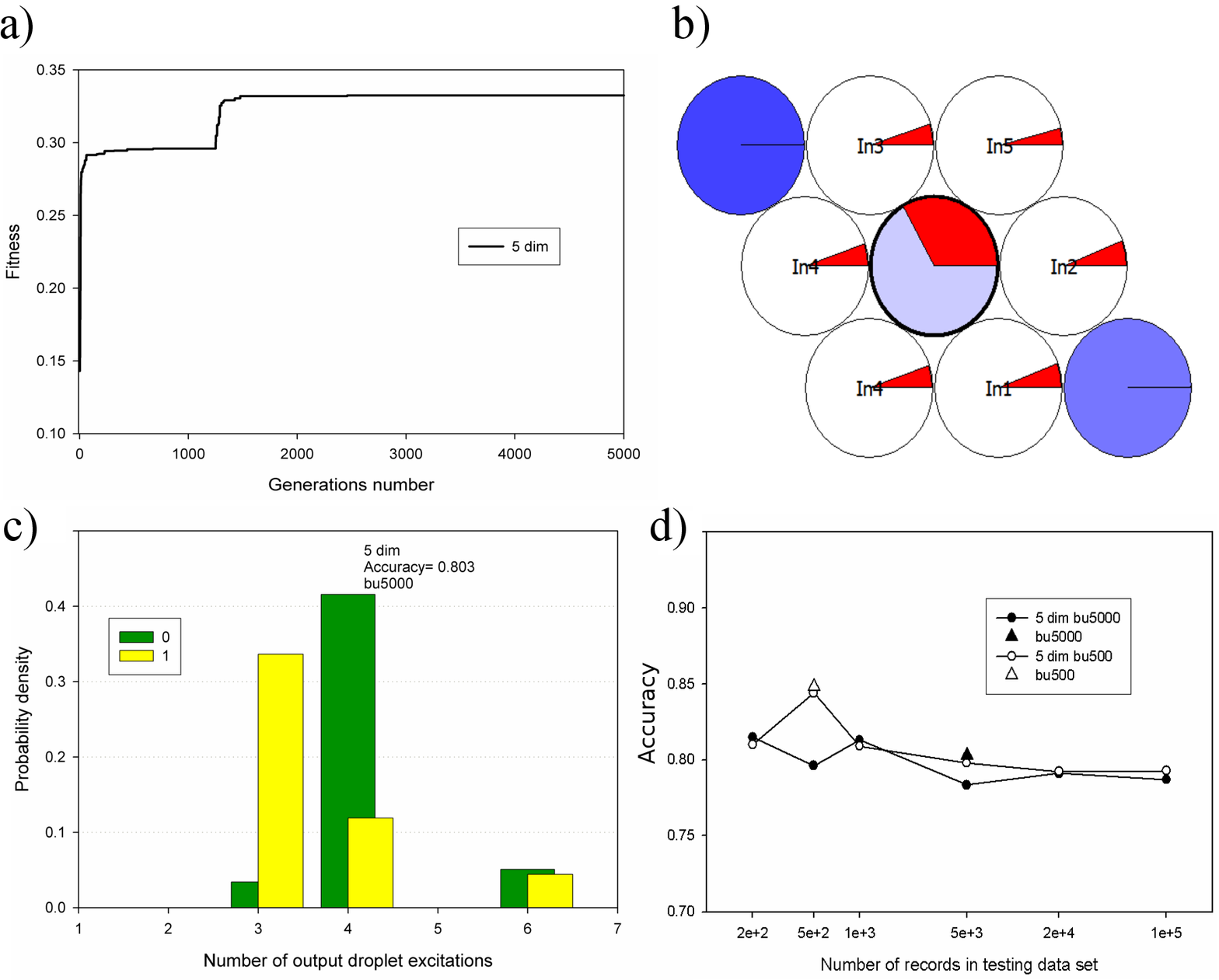}}
\end{center}
\caption{ The optimization of classifier for Ball-in-Cube problem in 5 dimensions. Notation as in Fig. 3. }
\label{fig-5}
\end{figure}

\clearpage

\begin{figure}
\begin{center}
\scalebox{0.7}{\includegraphics{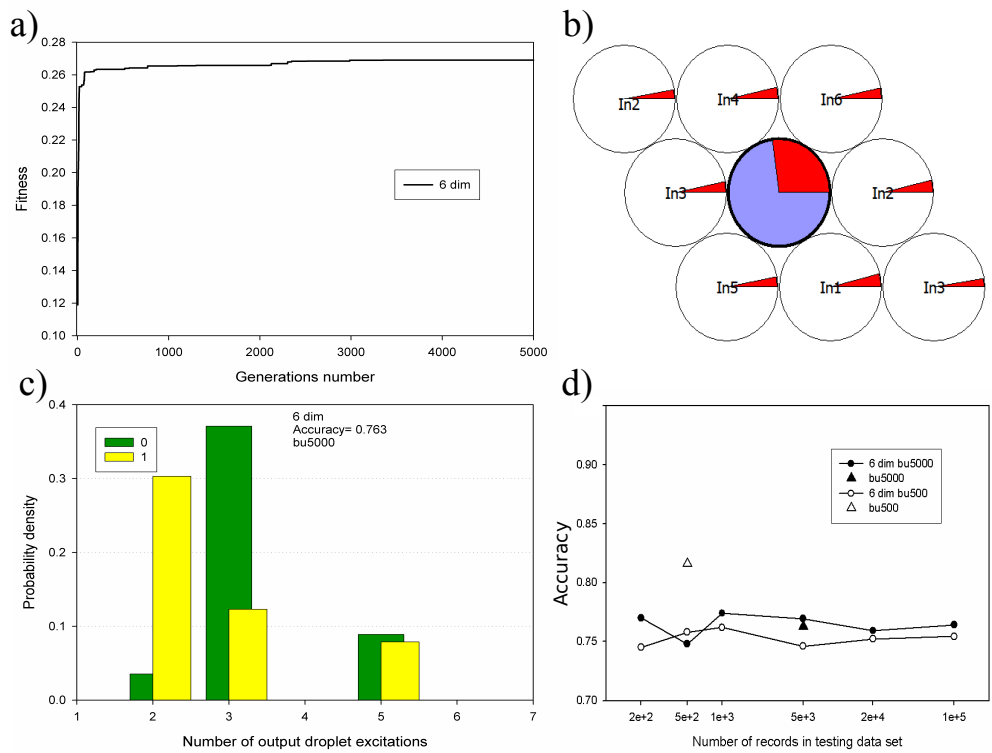}}
\end{center}
\caption{ The optimization of classifier for Ball-in-Cube problem in 6 dimensions. Notation as in Fig. 3. }
\label{fig-6}
\end{figure}

\clearpage

\section{Discussion and conclusions}

In the paper we studied predictive power of classifiers for the Ball-in-Cube problem for different dimensions of the space. We considered small networks composed of $9$ oscillators arranged on a hexagonal lattice with interactions between the nearest neighbors. They were used to classify datasets of the Ball-in-Cube problem in space dimensions from 3 to 6. In all cases the top-down design of a database classifier was successful. As shown in Fig. 7b after optimization of classifier parameters a small structure of droplets ( sufficient for 2 logic gates for information coded in pulses of excitation \cite{adamatz-PRE}), was able to recognize 
if a point belongs to the ball or not with accuracy from almost $90\%$ for 3 dimensions down to $76\% $ for the 6-dimensional problem. One can say that the optimized classifiers are still far from being perfect, but they are not expected to be. There is no proof that 9-oscillator classifier can accurately solve the Ball-in-Cube problem. Nevertheless, we expect that our optimization method can produce the best classifier of the problem that can be made using the considered medium.

We considered two optimization strategies with different sizes of training datasets (bu500 and bu5000). It was expected that the training dataset of bu5000 much better represents the problem than that of bu500. 
However, we have not observed a significant difference in results of both optimizations (cf. Fig. 8). For most of the cases the classifiers
obtained after longer optimization performed better on the test datasets, but the differences are not big. It indicates that for the considered problems even the small training datasets contained sufficient information about correlations between predictor values and the record type. The fact that for the considered optimization algorithms the number of generations does not much improve the classifier accuracy can be explained when we refer to the fitness as function of generation number illustrated in Figures 3-6. The increase in fitness is fast at the early stage of optimization. Next it remains constant and increases after rare fluctuations that reshape the classifier structure towards the higher maximum of the fitness.

The structures of classifiers optimized using bu500 and bu5000 strategies are compared in Fig.8. In all cases a normal droplet is the output one. Also in almost all cases (except of $n=4$ and bu5000) the input droplets of all predictors are the neighbors of the output droplet. Therefore, information about droplet symmetry contained in the training dataset was recognized by optimization algorithm and included into the classifier structure. It is worth to notice that in all classifiers of the inscribed ball problem in 3-dimensions, constructed using a regular network of droplets, the output droplet was the input one for the predictor $p_1$ \cite{cmst}, so the symmetry was broken. 

We limited our study to space dimensions not higher than 6.  In can be expected that the presented approach does not work at higher dimensions. The problem datasets is symmetrical with respect to all coordinates. In the considered network the output droplet can have maximum 6 directly interacting neighbors, thus for $n=7$ the input droplet for one of the coordinates will be excluded from the contact with the output and due to symmetry breaking the decrease in classifier accuracy can be expected. To achieve symmetry between inputs and the output for $n \ge 7$ a larger network should be concerned.

Despite the simplicity of the considered networks the optimized classifiers correctly reflect the correlations between predictor values and the record type characteristic for the problem. The tests on datasets of the problem containing up to 100 000 record have shown that even in the most difficult 6-dimensional case the optimized classifier perform its function with accuracy exceeding 75\%, whereas random answers should give 50\% of correct answers.
Therefore, it is justified to claim that the optimized classifiers do have predicting ability and, after a proper training, they can deal with record that were not included into the process of training. A similar conclusion followed from our previous study on 
the Wisconsin Breast Cancer Dataset \cite{cancer}. For this problem the optimized classifiers returned the correct answer in more than 80\% of cases not considered in the process of training.

The predictive ability is important for potential application of a classifier.
To construct a classifier we need chemical oscillators that interact with each other and can be controlled by an external factor. Many of such oscillators working at different conditions can be found. One field of applications includes smart drugs with a chemical classifiers built into 
their surrounding capsule. If there are warnings about infection in their neighborhood, then 
the classifier generates an output signal that opens the capsule and releases the drug. The results presented in this paper illustrate that a classifier working with a reasonable accuracy can be relatively small (if $n=6$ then $2/3$ of all droplets are the input ones). Moreover, even such simple classifiers, designed using small training datasets, can produce reasonably accurate answers. 

\clearpage

\begin{figure}
\begin{center}
\scalebox{0.6}{\includegraphics{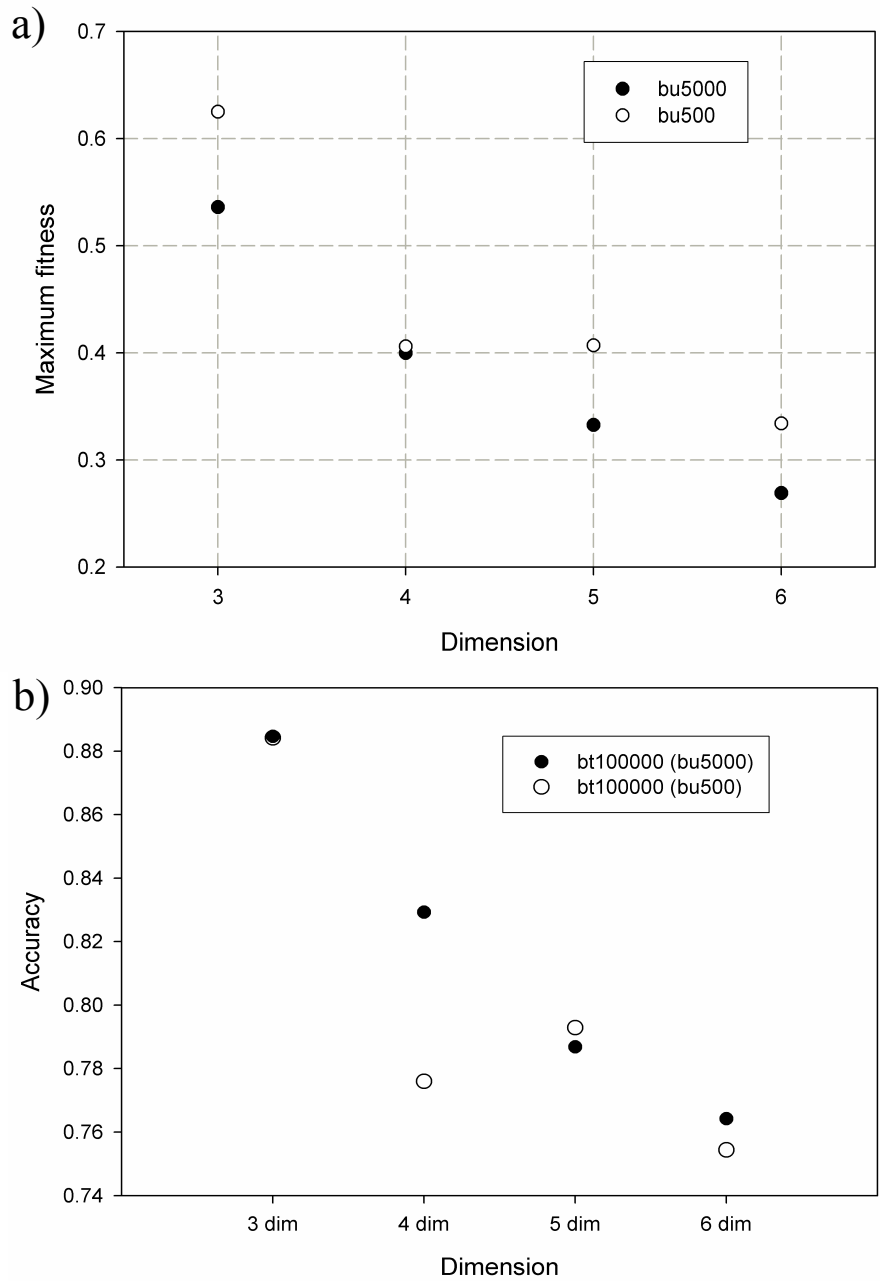}}
\end{center}
\caption{ Fitness (a) and accuracy (b) as the function of space dimension. The open symbols mark results obtained after a 
short optimization (bu 500), the filled ones are the results of a long optimization (bu5000). For $n=3$ the accuracy for bu500 is almost the same as for bu5000 (b).   } 
\label{fig-7}
\end{figure}

\clearpage

\begin{figure}
\begin{center}
\scalebox{0.6}{\includegraphics{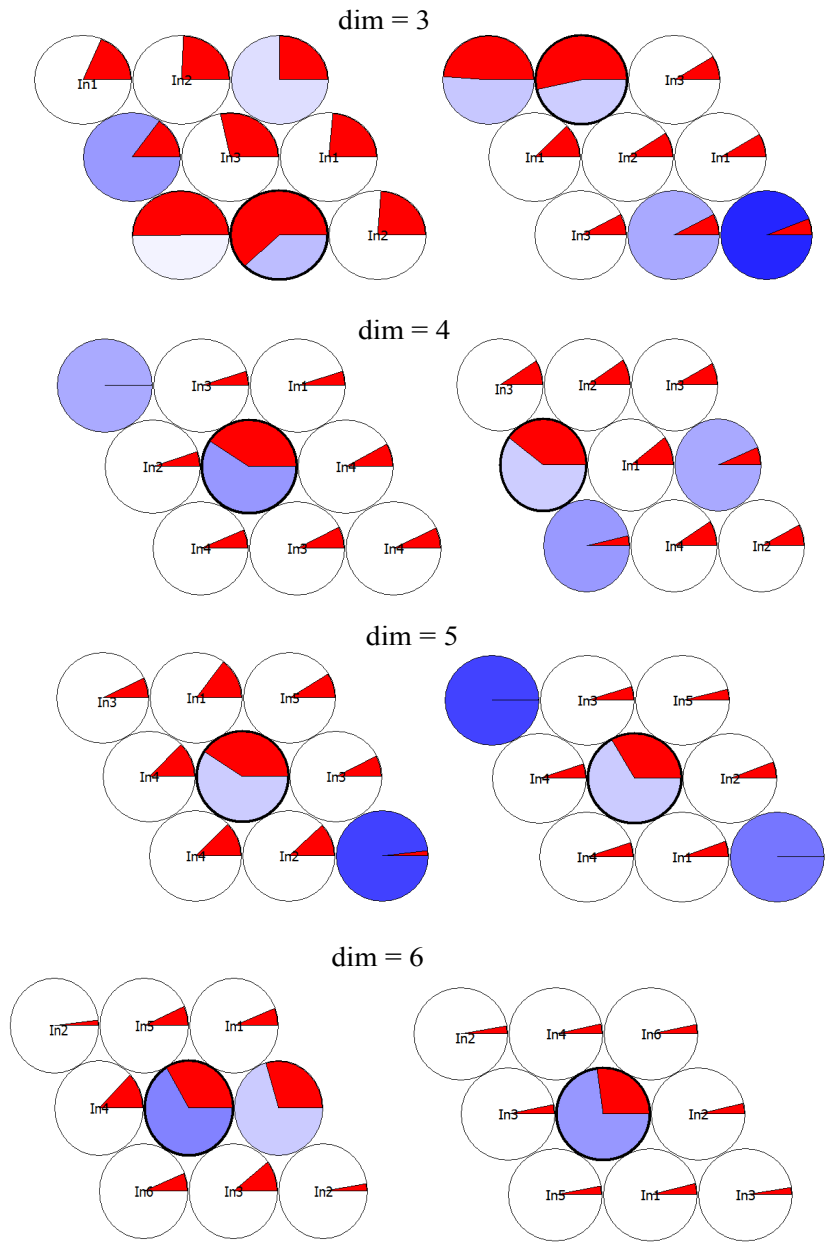}}
\end{center}
\caption{ Structures of classifiers obtained for training datasets containing $500$ records (left column) and $5000$ records (right column).
In the first case the optimization was performed for $500$ generations, in the second case $5000$ generations were used.
The fitnesses of illustrated classifiers were $\{0.625, 0.406, 0.407, 0.334\}$ and $\{0.536, 0.400, 0.333, 0.269\}$ for space dimensions from 3 to 6.
The accuracy of training base classification were $\{0.9220, 0.7960, 0.8480 , 0.8160\}$ and $\{0.8942, 0.8420, 0.8028, 0.7628\}$ for space dimensions from 3 to 6.
The arc length of red slice, normalized to the circumference gives the mutual information between the number of droplet excitations and record types measured in bits.}
\label{fig-8}
\end{figure}

\clearpage

\section{Acknowledgment}
The work was supported by the Polish National Science Centre grant UMO-2014/15/B/ST4/04954.

\end{document}